\documentclass[aip,rsi,amsmath,amssymb,reprint]{revtex4-1}
\usepackage{graphicx}
\usepackage{dcolumn}
\usepackage{bm}
\usepackage{hyperref}

\draft 

\begin{document}





\title{A compact and efficient strontium oven for laser-cooling experiments} 








\author{M. Schioppo}
\affiliation{Dipartimento di Fisica e Astronomia and LENS, Universit\`a di Firenze and INFN Sezione di Firenze, \\
Via Sansone 1, 50019 Sesto Fiorentino, Italy}

\author{N. Poli}
\affiliation{Dipartimento di Fisica e Astronomia and LENS, Universit\`a di Firenze and INFN Sezione di Firenze, \\
Via Sansone 1, 50019 Sesto Fiorentino, Italy}

\author{M. Prevedelli}
\altaffiliation[Permanent address: ]{Dipartimento di Fisica, Universit\`a di Bologna, Via Irnerio 46, 40126 Bologna, Italy}
\affiliation{Dipartimento di Fisica e Astronomia and LENS, Universit\`a di Firenze and INFN Sezione di Firenze, \\
Via Sansone 1, 50019 Sesto Fiorentino, Italy}

\author{St. Falke}
\affiliation{Physikalisch-Technische Bundesanstalt, Bundesallee 100, 38116 Braunschweig,Germany}

\author{Ch. Lisdat}
\affiliation{Physikalisch-Technische Bundesanstalt, Bundesallee 100, 38116 Braunschweig,Germany}

\author{U. Sterr}
\affiliation{Physikalisch-Technische Bundesanstalt, Bundesallee 100, 38116 Braunschweig,Germany}

\author{G.M. Tino}
 \email{Guglielmo.Tino@fi.infn.it}
\affiliation{Dipartimento di Fisica e Astronomia and LENS, Universit\`a di Firenze and INFN Sezione di Firenze, \\
Via Sansone 1, 50019 Sesto Fiorentino, Italy}





\date{\today}

\begin{abstract}

Here we describe a compact and efficient strontium oven well suited for laser-cooling experiments. Novel design solutions allowed us to produce a collimated strontium atomic beam with a flux of $1.0\times10^{13}\,\text{s}^{-1}\text{cm}^{-2}$ at the oven temperature of $450\,^{\circ}\text{C}$, reached with an electrical power consumption of $36\,$W. The oven is based on a stainless-steel reservoir, filled with $6\,$g of metallic strontium, electrically heated in a vacuum environment by a tantalum wire threaded through an alumina multi-bore tube. The oven can be hosted in a standard DN40CF cube and has an estimated continuous operation lifetime of 10 years. This oven can be used for other alkali and alkaline earth metals with essentially no modifications.
\end{abstract}

\pacs{}

\maketitle 



\section{Introduction}
Laser cooling is today an established technique to produce samples of atoms with temperature approaching the zero Kelvin limit. Ultra-cold strontium samples are largely studied, with experiments ranging from quantum degeneracy, quantum computation, to applications as quantum sensors of force and highly accurate optical clocks. These experiments are all based on an oven, which provides the needed flux of atoms in an ultra high vacuum (UHV) environment (pressure $<10^{-9}\,$mbar). At room temperature the vapor pressure of alkaline earth metals (as strontium) is lower than that of alkali at the same temperature, for this reason relatively high temperature ovens are typically needed ($T\gtrsim450^{\circ}$C for $\text{Sr}$). In particular the application of alkaline earth metals for optical clocks is today prompting significant scientific and technological efforts for the realization of compact, reliable and transportable apparatus, in the perspective of a their future use in space \cite{Tino:2007, Schiller:2007,Schiller:2009, SOC:2012}. This background is at the basis of our study on an efficient strontium oven, which represents one of the critical parts in an optical clock in terms of power consumption, size and heat management. 

Here we describe a compact and efficient strontium oven essentially based on a heater placed in vacuum. The novelty of the proposed oven concerns the synthesis of the solutions proposed so far for high efficient atomic beam sources (for a general review of atomic sources see \cite{Ross:1995}), such as capillaries and thermal shield, into a compact design, for both horizontal and vertical operation, without the complication of independent heaters for capillaries and crucible and without water cooling of the thermal shield. Additionally reduced size and high efficiency (high flux at lower temperature and power consumption) imply a simplified heat management of the test chamber where clock spectroscopy is performed, with important consequences on the uncertainty budget of \text{Sr} clock measurement, nowadays limited by the lack of control of the homogeneity of the blackbody environment \cite{Campbell:2008, Falke:2011,Yamaguchi:2012jr}. The use of this oven can be extended to other alkali and alkaline earth metals not reacting with stainless-steel at high temperature.

\section{Oven design}
\begin{figure}
\includegraphics{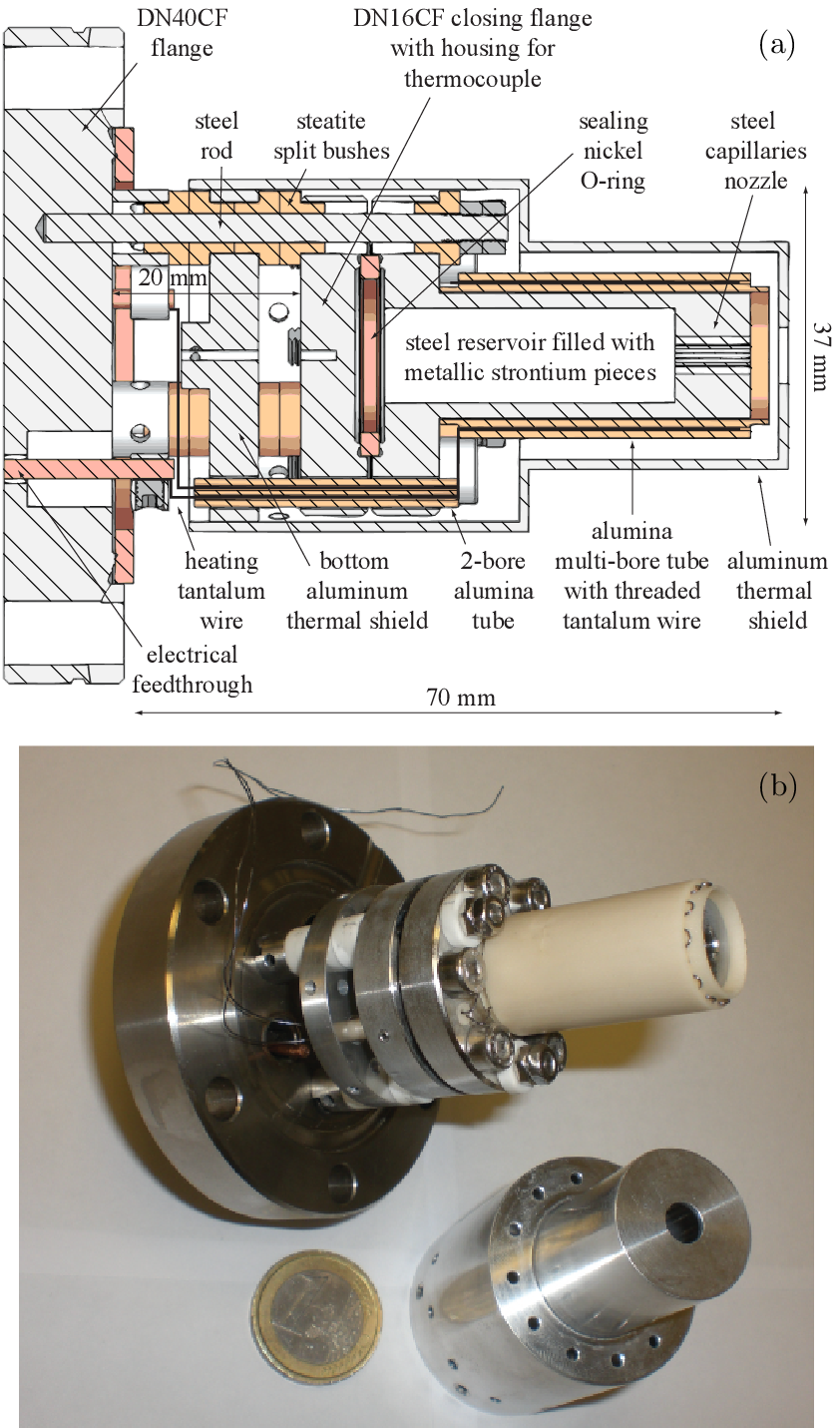}
\caption{\label{fig:oven_design} (Color online) (a) Section view of the compact strontium oven showing the internal components and layout. (b) Picture of the oven with the aluminum thermal shield. For the sake of comparison a one euro coin is displayed.}
\end{figure}
The oven design is shown in Fig. \ref{fig:oven_design}. About 6$\,$g of metallic strontium are placed in a stainless-steel reservoir, with internal length and diameter respectively of 34$\,$mm and 10$\,$mm. The reservoir is electrically heated in vacuum by a 0.3 mm diameter tantalum wire threaded through an alumina (AL23) multi-bore tube. This material has been chosen since it ensures both a high electrical insulation (specific resistance $10^{10}\;\Omega\,\text{cm}$ at $500\,^{\circ}\text{C}$) and a good thermal conductivity ($11\;\text{W}\,\text{m}^{-1}\text{K}^{-1}$ at $500\,^{\circ}\text{C}$). The reservoir is closed by a DN16CF flange (removable for strontium refilling) sealed by a standard nickel gasket. The opposite side of the reservoir is terminated by a nozzle (diameter $4\,$mm, length $8\,$mm) filled with $N_{\text{cap}}\simeq120$ stainless-steel capillaries $L=8\,$mm long and with an internal radius $a=100\,\mu\text{m}$. The high temperature assembly ``reservoir + closing flange" is supported by three stainless-steel rods (diameter $3\,$ mm) welded on a DN40CF flange. The thermal contact between the high temperature assembly and the supporting rods is reduced by using steatite split bushes UHV cleaned, having a thermal conductivity of $2.5\;\text{W}\,\text{m}^{-1}\text{K}^{-1}$ at $500\,^{\circ}\text{C}$, about one order of magnitude below that of 304 stainless-steel at the same temperature. The oven can be hosted in a standard DN40CF cube, which can be connected to an ion pump and to a generic vacuum system for laser-cooling experiments. The vacuum environment prevents the convection mechanism of heat transfer. The remaining heat transfer channel is represented by the thermal radiation (blackbody radiation) of the hot reservoir at $450\,^{\circ}\text{C}$. In order to shield this radiation the oven is surrounded by a $1\,$mm thick aluminum cylinder. Aluminum has been chosen for its low emissivity coefficient $\varepsilon\simeq0.2$.

Electrical power is applied by means of two UHV electrical feedthroughs. Here two stainless-steel clamps are used to electrically connect the two feedthrough pins to the tantalum heating wire. Short circuits are prevented by beading the two tantalum wire terminations with alumina tubes (0.4 mm internal diameter). Tantalum wires go through the DN16CF flange and reservoir and finally into the multi-bore tube heater by means of an alumina tube 28 mm long and with two $0.8\,$mm diameter holes.

The temperature of the oven is measured by a thermocouple in thermal contact with the rear side of the DN16CF flange. An aluminum plate is placed between the DN40CF and the DN16CF flange to shield the blackbody radiation. The thermocouple is electrically connected through the DN40CF flange by means of an UHV feedthrough.       

\section{Experimental results}

\subsection{Thermal properties}
\begin{figure}[]
\includegraphics{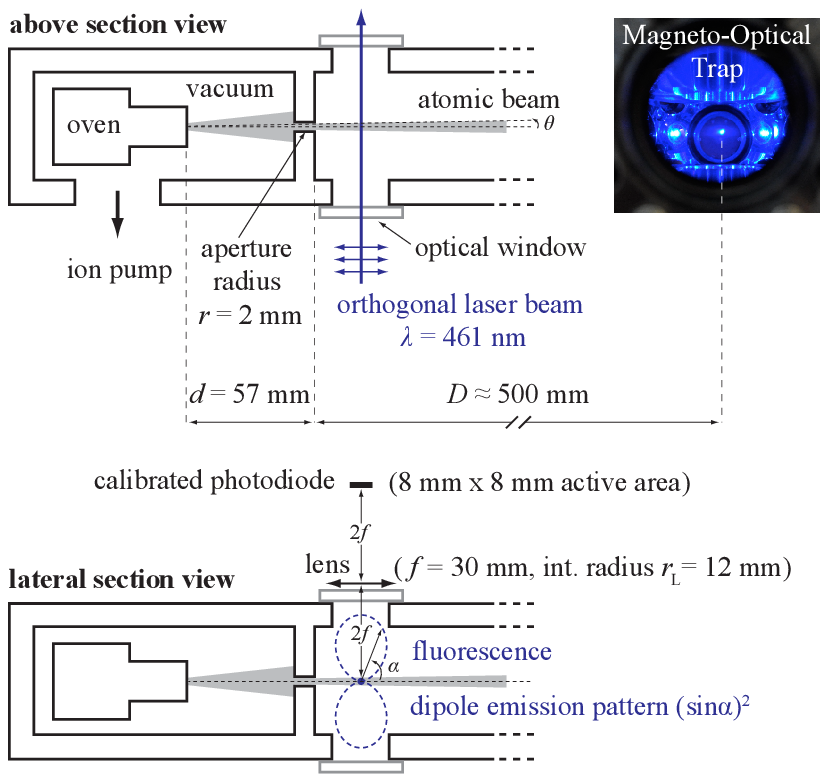}
\caption{\label{fig:experimental_setup} Schematic view of the experimental setup. The oven is placed in vacuum and the atomic beam is further collimated by an aperture with radius $r$ placed at distances $d$ and $D$ respectively from the nozzle and magneto-optical trapped atoms. An orthogonal laser beam resonant with the strontium transition at $\lambda=461\,$nm and horizontally polarized is used to estimate flux and divergence of the atomic beam by measuring the atomic fluorescence.}
\end{figure}
\begin{figure}[]
\includegraphics{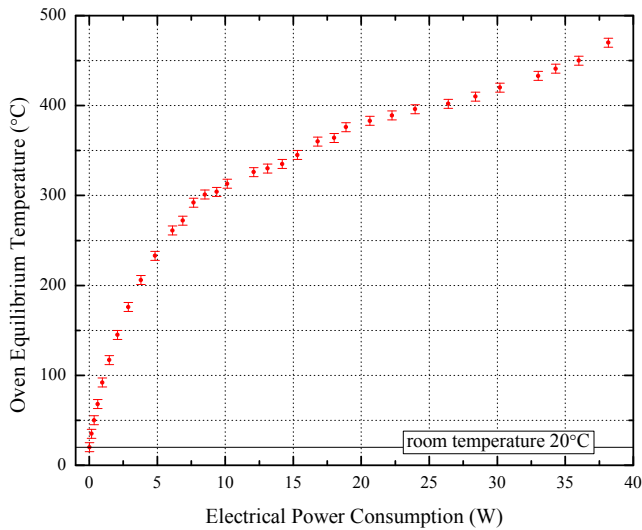}
\caption{\label{fig:thermal_properties} Oven equilibrium temperature as a function of the electrical power consumption. The oven working temperature of $450\,^{\circ}$C is reached with $36\,$W of power consumption. The vertical error bars are given by the temperature uncertainty of the thermocouple ($\pm\,5\,^{\circ}$C).}
\end{figure}

In our experimental setup (see Fig. \ref{fig:experimental_setup}) the oven is placed in a custom vacuum chamber evacuated by a $40\,$l/s ion pump. An aperture with radius $r=2\,$mm is placed at a distance $d=57\,$mm from the nozzle of the oven to further increase the collimation of the atomic beam ($r/d\simeq35\,$mrad) and to avoid the effusion of strontium toward the optical windows of the vacuum chamber. The equilibrium pressure in the vacuum chamber at the working oven temperature of $450\,^{\circ}$C is $\sim10^{-8}\,$mbar. In such environment this temperature is reached with a power consumption of $36\,$W ($2.38\,$A - $15.2\,$V). Fig. \ref{fig:thermal_properties} shows that the measured equilibrium temperature $T$ is not linear with the electrical power consumption $\text{P}$, since the thermal properties of the used materials depend on temperature and the loss of energy by thermal radiation increases with $T^4$. However it is possible to define a total thermal resistance $R_{\text{th}}(T)$ given by  
\begin{equation}
T-T_{\text{env}}=R_{\text{th}}(T)\times\text{P}\,
\label{eq:thermal_resistance}
\end{equation}
where $T_{\text{env}}=20\,^{\circ}\text{C}$ is the room temperature. At the working temperature of $450\,{^\circ}\text{C}$ we experimentally find a thermal resistance $R_{\text{th}}\simeq12\,^{\circ}\text{C}/\text{W} $. 

Maximum temperature is limited by the thermal shield in aluminum having the melting point at $660\,{}^{\circ}$C. However by using stainless-steel it is possible to increase the maximum operating temperature at the $\sim1000\,{}^{\circ}$C level.

\begin{figure}[]
\includegraphics{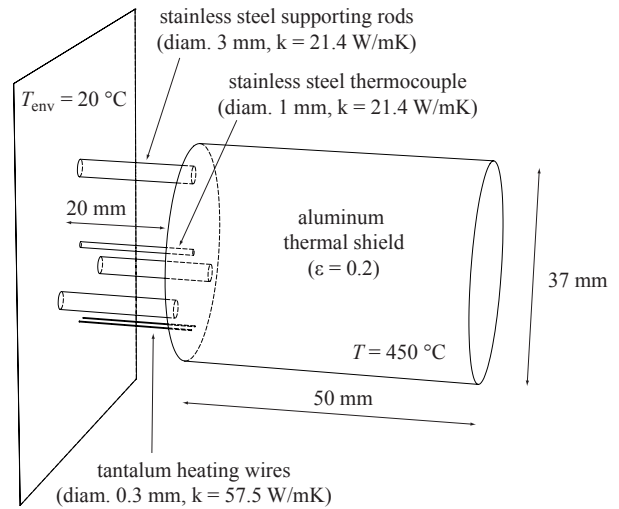}
\caption{\label{fig:thermal_model} Simplified thermal model of the proposed oven to trace out the origin and nature of the heat losses (see the text).}
\end{figure}

In order to evaluate whether the main heat losses are due to conduction or blackbody radiation (at $450\,^{\circ}$C) we consider a simplified thermal model of the oven (as shown in Fig. \ref{fig:thermal_model}) with the aluminum thermal shield at $450\,^{\circ}$C in contact with the environment at $20\,^{\circ}$C (at  distance of $20\,$mm) through three stainless-steel supporting rods, one stainless-steel thermocouple and two tantalum heating wires. From the simple expression $\text{P}_{\text{cond}}=kA\Delta T/l $, where $k$ is the thermal conductivity, $A$ is the area of the conducting surface, $\Delta T=T-T_{\text{env}}$ is the temperature difference and $l$ is the thickness of the conducting surface separating the two temperatures, we find that the heat conduction loss due to the supporting rods is $10\,$W, thermocouple and heating wires together are at level of $0.5\,$W. The main loss is due to the blackbody radiation emitted by the aluminum thermal shield at $450\,^{\circ}$C, given by
\begin{equation}
\text{P}_{\text{rad}}=S_{\text{tot}}\sigma\varepsilon\left(T^{4}-T_{\text{env}}^{4}\right)\simeq24\,\text{W}\,
\label{eq:BBR:loss}
\end{equation} 
where $S_{\text{tot}}=80\,\text{cm}^2$ is total surface area of the thermal shield, $\sigma$ is the Stefan-Boltzmann constant, and $\varepsilon\simeq0.2$ is the emissivity of aluminum. The total estimated power loss is below $35\,$W, close to the measured value $36\,$W. The latter value can be now put in perspective and we can conclude that the thermal properties of the oven can be further improved by choosing for the thermal shield a material or a surface finishing (lapping or plating) with reduced emissivity ($\varepsilon<0.2$). 

Additionally, compact size and high efficiency of the oven ensure a better control on the blackbody radiation (BBR) coming from the hot furnace into the atoms trapping region. In our experimental configuration atoms are magneto-optical trapped at a distance $D\simeq500\,$mm from the cold (at room temperature) aperture, leading to a total BBR frequency shift due to the oven, on the strontium clock transition $5\text{s}^2\,{}^1\text{S}_0-5\text{s}5\text{p}\,{}^3\text{P}_0$, given in fractional units by \cite{Porsev:2006, Middelmann:2011}

\begin{equation}
\left|\frac{\delta\nu_{\text{BBR}}^{\text{(oven)}}}{\nu_{0}}\right|=5.484\cdot10^{-15}\times\left(\frac{T}{300\,\text{K}}\right)^{4}\times\frac{\Theta}{4\pi}<10^{-18}\, 
\label{eq:BBR}
\end{equation}
where $\nu_0\simeq429\,\text{THz}$ is the frequency of the strontium clock transition, $T=723\,\text{K}\,$ is the temperature of the oven and $\Theta=\pi(r/D)^2$ is the solid angle under which the atoms see the cold aperture. Remarkably this estimation makes the BBR frequency shift produced by the oven in this configuration completely negligible in the state of art strontium clocks corrections and uncertainty budget \cite{Campbell:2008, Falke:2011,Yamaguchi:2012jr}. The latter is nowadays dominated by the BBR effect of the environment at room temperature and a compact and thermally insulated oven can additionally simplify the heat management of the test chamber (where the clock spectroscopy is performed) and improve its temperature homogeneity with a great benefit in the final uncertainty budget of strontium clock.

\subsection{Orthogonal crossed-beam spectroscopy for laser frequency stabilization}
\begin{figure}[]
\includegraphics{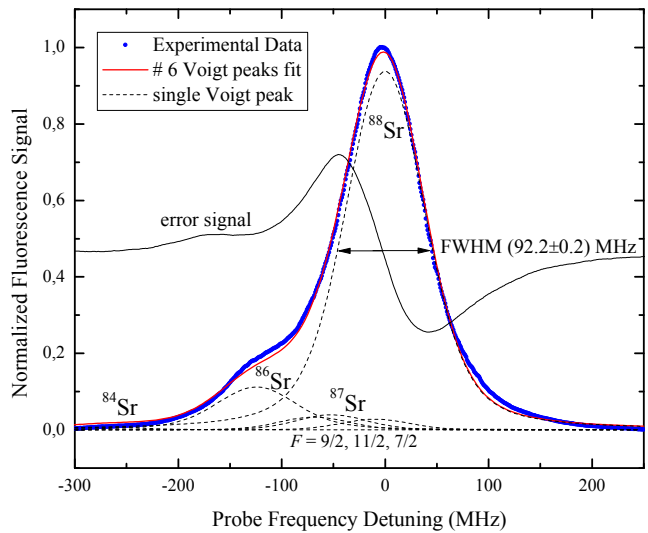}
\caption{\label{fig:single_beam_spectroscopy} (Color online) Fluorescence spectrum of the strontium $5\text{s}^2\,{}^1\text{S}_0\leftrightarrow5\text{s}5\text{p}\,{}^1\text{P}_1$ transition (blue dots) obtained in an orthogonal crossed-beam spectroscopy with a single probe beam. The fit function (red line) given by the sum of six Voigt functions (adjusted r-square of fit 0.998) is used to calibrate the frequency axis. Each Voigt peak (black dashed lines) represents the contribution of the three bosonic isotopes $^{84}\text{Sr}$, $^{86}\text{Sr}$, $^{88}\text{Sr}$ and the three hyperfine components of the fermionic $^{87}\text{Sr}$. The error signal associated to the spectrum can be used to frequency stabilize the laser on the atomic transition.}
\end{figure}

In order to characterize the atomic beam produced by the oven, spectroscopy on the strontium transition $5\text{s}^2\,{}^1\text{S}_0\leftrightarrow5\text{s}5\text{p}\,{}^1\text{P}_1$ at $\lambda=461\,$nm (saturation intensity $I_s=42.7\,\text{mW/}\text{cm}^2$) is performed. A linearly polarized probe beam, with a diameter  $\phi=2.35\,$mm ($1/e^2$) and an optical power of $165\,\mu$W (saturation parameter $s\equiv I/I_s\simeq0.09$), is sent orthogonally with respect to the direction of the atomic beam (see Fig. \ref{fig:experimental_setup}). In this linear absorption regime with a single probe beam, the atomic fluorescence signal is measured as a function of the probe frequency (see Fig. \ref{fig:single_beam_spectroscopy}). This spectrum is fitted with the sum of six Voigt functions, one for each bosonic isotope ($^{84}\text{Sr}$, $^{86}\text{Sr}$, $^{88}\text{Sr}$) and hyperfine components of the fermionic $^{87}\text{Sr}$  ($F=9/2,\,11/2,\,7/2$). By knowing the natural linewidth ($\gamma=32\,\text{MHz}$), the isotope shifts, hyperfine splittings, relative strengths and the natural abundances \cite{Mauger:2007}, it is possible to obtain from the fit the remaining five free parameters given by an overall amplitude and offset, a center frequency offset, the gaussian width (not calibrated) and more importantly the time/frequency scaling factor to calibrate in frequency our scan. With this calibration we obtain a full-width-half-maximum of $(92.2\pm0.2)\,\text{MHz}$ for the $^{88}\text{Sr}$ peak. This value essentially depends on the transverse velocity distribution of atoms (or equivalently to the collimation of the atomic beam). Indeed, the limited transit-time of atoms across the probe beam $\Delta t=\phi/v_{\text{beam}}$, with $v_{\text{beam}}$ the most probable longitudinal speed in an effusive atomic beam
\begin{equation}
v_{\text{beam}}=\sqrt{3k_{\text{B}}T/M}\simeq450\,\text{m/s}\quad(T=450\,^{\circ}\text{C})
\label{max_longitudinal_velocity}
\end{equation}
($k_\text{B}$ Boltzmann constant, $M$ strontium mass), corresponds to a Fourier-limited frequency width $1/\Delta t$ of $190\,\text{kHz}$, that can be neglected. Now considering for simplicity just one isotope peak, by assuming a normal transverse velocity distribution
\begin{equation}
f_{t}(v_{t})\propto\exp\left(-\frac{v_{t}^{2}}{2\sigma_{t}^{2}}\right) 
\label{eq:transverse_distribution}
\end{equation} 
where $\sigma_t$ is the velocity standard deviation, the fluorescence signal is given by the Voigt profile
\begin{equation}
F(\delta\omega)\propto\int_{-\infty}^{+\infty}\frac{f_{t}(v_{t})}{\left(\delta\omega-\left|\mathbf{k}\right|v_{t}\right)^{2}+\Gamma^{2}/4}\, dv_{t} 
\end{equation}
where $\delta\omega=\omega-\omega_0$ is the angular frequency detuning from the resonance $\omega_0$, $\left|\mathbf{k}\right| =2\pi/\lambda$ is the modulus of the probe beam wavevector (here we assumed a perfect orthogonality) and $\Gamma=2\pi\gamma$ is the natural linewidth (in angular frequency). We find from the frequency axis calibration that the fit value of the standard deviation $\sigma_t$ is $(14.41\pm0.04)\,$m/s, from which we obtain the experimental value of the atomic beam collimation 
\begin{equation}
\theta\equiv\frac{\sigma_{t}\sqrt{2\ln2}}{v_{\text{beam}}}\simeq37\,\text{mrad} 
\label{eq:collimation}
\end{equation}
where $\sigma_{t}\sqrt{2\ln2}$ is the half width half maximum of the transverse velocity distribution in Eq. (\ref{eq:transverse_distribution}). The measured value of the beam collimation is very close to the angular selection given by the presence of the aperture ($r/d\simeq35\,$mrad).

Such collimated atomic beam is a good tool to frequency stabilize a laser at the MHz level on the strontium transition $5\text{s}^2\,{}^1\text{S}_0\leftrightarrow5\text{s}5\text{p}\,{}^1\text{P}_1$ (error-signal in Fig. \ref{fig:single_beam_spectroscopy}) and is thus an alternative to hollow cathode lamps, sealed vapor cells \cite{Tino:1992, Bridge:2009} and split-photodiode technique \cite{Jitschin:1984}. As the transverse Doppler width is comparable to the natural linewidth, the Doppler-free saturation spectroscopy on the atomic beam has not advantage with respect to single beam excitation. For this application the orthogonality between the probe and the atomic beam is critical to avoid any systematic shift $\delta\nu$ between the atomic resonance and the stabilized laser frequency. The sensitivity to this effect is given by the projection of the most probable longitudinal velocity onto the probing direction 
\begin{equation}
\delta\nu\sim\frac{v_{\text{beam}}}{\lambda}\varphi\simeq0.98\,\text{MHz/mrad} 
\label{eq:angular_sensitivity}
\end{equation}  
where $\varphi\,$($\ll1\,\text{rad}$) represents the angular deviation from the perpendicular to the atomic beam. The condition of orthogonality (and zero systematic frequency shift) is experimentally achieved by maximizing the peak and minimizing the width of the fluorescence signal. Additionally an estimation of the residual systematic shift $\delta\nu$ can be obtained from the width variation when the probe beam is retroreflected and is given by the simple expression 
\begin{equation}
\delta\nu\simeq\left(\Delta\nu_{\text{refl.}}-\Delta\nu_{\text{single}}\right)/2 
\end{equation}
where $\Delta\nu_{\text{refl.}}$ and $\Delta\nu_{\text{single}}$ represents respectively the width of the fluorescence signal with and without retro-reflection. In our experimental configuration we find a systematic shift (modulus) lower than $2\,$MHz, well below the natural linewidth of the strontium $5\text{s}^2\,{}^1\text{S}_0\leftrightarrow5\text{s}5\text{p}\,{}^1\text{P}_1$ transition ($\gamma=32\,$MHz). 

\subsection{Atomic beam flux, effusion regime and duration}
In order to estimate the atomic flux, the fluorescence peak produced in the orthogonal spectroscopy is measured by means of a calibrated photodiode. The flux of atoms $\mathcal{F}$ ($\text{atoms}\,\text{s}^{-1}\,\text{cm}^{-2}$) is given by
\begin{equation}
\mathcal{F}=\rho\, v_{\text{beam}} 
\end{equation} 
with $\rho$ atomic density. The latter is related to the measured fluorescence optical peak power $\mathcal{P}_{\text{max}}$ (W) through
\begin{eqnarray}
\rho=&&\frac{1}{V_{\text{int}}}\times\mathcal{P}_{\text{max}}\times\frac{\Omega_{\text{tot}}}{\Omega_{\text{ph}}} \nonumber\\
&&
\times\left(\frac{\hslash\omega_{0}\Gamma s}{2\sigma_t\sqrt{2\pi}}\int_{-\infty}^{+\infty}\frac{\exp\left(-v_{t}^{2}/2\sigma_{t}^{2}\right)}{1+4\left(\left|\mathbf{k}\right|v_{t}/\Gamma\right)^{2}}dv_{t}\right)^{-1}
\end{eqnarray}
where $V_{\text{int}}\approx\pi(\phi/2)^2 2r$ is the interaction volume, $\Omega_{\text{tot}}=\int_{0}^{\pi}(\sin\alpha)^{2}(2\pi\sin\alpha\,d\alpha)=8\pi/3$ is total solid angle of dipole emission, $\Omega_{\text{ph}}=\pi(r_{\text{L}}/2f)^{2} $ is the solid angle under which the atomic fluorescence is seen by the photodiode (see Fig. \ref{fig:experimental_setup}) and $\hslash$ is the reduced Planck's constant. At $T=450\,^{\circ}\text{C}$ taking into account the contributions of all the strontium isotopes we estimate a total flux of $\mathcal{F}\simeq1.0\times10^{13}\,\text{s}^{-1}\text{cm}^{-2}$, corresponding to a flow rate of $\dot{N}=\mathcal{F}(\pi r^2)\simeq1.2\times10^{12}\,\text{s}^{-1}$. Since from Eq. \ref{eq:collimation} we find that atomic beam divergence ($\theta \simeq 37\,\text{mrad}$) is determined by the cold aperture ($r/d\simeq35\,\text{mrad}$) and not by the geometry of the capillaries ($\theta_{\text{t}}\equiv1.68\,a/L\simeq21\,\text{mrad}$ \cite{Giordmaine:1960}), we can define an atomic beam intensity peak ($\text{atoms}\,\text{s}^{-1}\,\text{steradian}^{-1}$) given by
\begin{equation}
J\equiv\frac{\dot{N}}{\Phi} 
\end{equation}
where $\Phi=\pi(r/d)^2$ is the solid angle selection performed by the cold aperture (see Fig. \ref{fig:experimental_setup}). In order to establish the effusion regime the atomic beam intensity peak is measured as a function of the oven temperature and is compared with the theoretical calculation in the limit of no collisions occurring in the capillaries (Knudsen regime) \cite{Giordmaine:1960, Beijerinck:1975}
\begin{equation}
J_{\text{t}}=\frac{(\pi a^{2})\bar{v}n}{4\pi}N_{\text{cap}} 
\end{equation}
where $\bar{v}=\left(8k_{\text{B}}T/\pi M\right)^{1/2}$ and $n=P/k_{\text{B}}T$ are respectively the mean atomic speed and the atomic density in the reservoir, with $P$ vapor pressure of strontium given by \cite{CRC}
\begin{equation}
\log\left(P(\text{Pa})\right)=14.232-\frac{8572}{T(\text{K})}-1.1926\log\left(T(\text{K})\right)\,. 
\end{equation}
\begin{figure}[]
\includegraphics{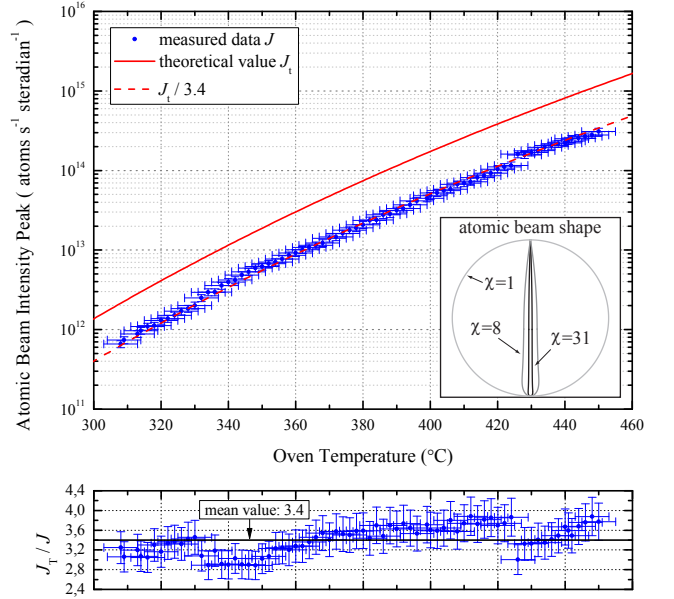}
\caption{\label{fig:atomic_flux} Semilog plot of the measured atomic beam peak intensity (dots) as a function of the oven temperature, in comparison with the theoretical curve (solid curve) for a collision-free source. Experimental and theoretical values are in agreement except for a proportional factor of $\sim3.4$ (see the text). The dashed line represents the theoretical value scaled by this factor. Horizontal error bars of $\pm\,5\,^{\circ}\,$C are given by the temperature uncertainty of the thermocouple, vertical error bars of $10\,\%$ take into account the total uncertainty on the atomic beam intensity measurement (due to the calibration of the photodiode, temperature and geometrical factors). The inset shows the polar plot of the atomic beam intensity as a function of the emission angle \cite{Beijerinck:1975} for different values of the peaking factor $\chi$.}
\end{figure}
From the comparison, shown in Fig. \ref{fig:atomic_flux}, we find that the behavior of the experimental atomic intensity as a function of the oven temperature is in good agreement with the theoretical curve, except for a proportional factor of $\sim3.4$. We conclude that the atomic source is operating in the collision-free regime, as also confirmed by the calculation of the mean free path $l$ of strontium atoms inside the reservoir  (at $T=450\,^\circ$C)
\begin{equation}
l=\left(\sqrt{2}\pi n\sigma^{2}\right)^{-1}\simeq13\,\text{cm}\quad\gg\quad a\,,\, L 
\end{equation}
with $\sigma=4.3\,$\r{A} the atomic diameter of strontium. The disagreement between the experimental and theoretical absolute values of the atomic beam intensity, together with the evidence that the measured atomic beam divergence is larger than the theoretical Knudsen regime value $\theta_{\text{t}}$ \cite{Giordmaine:1960} can be both explained by a slightly imperfect mutual alignment of the capillaries. To understand more quantitatively if the reduced intensity is compliant with the measured increased divergence we can assume that for a small misalignment of the capillaries the number of atoms effused in the half-intensity solid angle $J\pi\theta^2$ is the same of ideal case $J_{\text{t}}\pi\theta_{\text{t}}^{2}$, leading to $J_{\text{t}}/J\sim(\theta/\theta_{\text{t}})^{2}\simeq3$, which is very close to the measured value.

In our collision-free regime the total flow rate of atoms emitted in the half solid angle by the oven ($T=450\,^{\circ}$C)  is given by \cite{Giordmaine:1960}  
\begin{equation}
\dot{N}_{\text{tot}}=\frac{2\pi}{3}\frac{n\bar{v}a^{3}}{L}N_{\text{cap}}\simeq1.2\times10^{14}\,\text{s}^{-1} 
\end{equation}
corresponding  to a continuous operation lifetime (with $6\,$g of strontium) of about $10\,$years. The system started continuous operation about 2 years ago, it is presently running without appreciable degradation of the atomic flux and it is regularly used for loading $\sim10^8$ atoms in a magneto-optical trap operating on the $5\text{s}^2\,{}^1\text{S}_0\leftrightarrow5\text{s}5\text{p}\,{}^1\text{P}_1$ transition.

Considering that the measured value of the peak intensity at $450\,^{\circ}$C is $J\simeq3.1\times10^{14}\,\text{s}^{-1}\,\text{steradian}^{-1}$, we can evaluate the peaking factor \cite{Beijerinck:1975} of our source
\begin{equation}
\chi\equiv\pi J/\dot{N}_{\text{tot}}\simeq8 
\end{equation}
where $\pi$ is added so that $\chi=1$ for an effusive cosine emitter source (ideal thin-walled orifice). The measured $\chi$ has to be compared with the theoretical value $\chi_{\text{t}}\equiv1/W=31$ (see inset of Fig. \ref{fig:atomic_flux}), where $W=(8a/3L)/(1+8a/3L)$ is the transmission probability (or Clausing factor) of the capillaries. We can conclude that because of the difficulty in aligning 120 capillaries with $L/a=80$ the theoretical peaking factor $\chi_{\text{t}}=31$ is not reached. The measured value $\chi\simeq8$ can be realized by employing a significantly simplified capillaries nozzle with $L/a\simeq20$.

\section{Conclusion}
In summary, we have presented a compact (length $70\,$mm, diameter $37\,$mm) and efficient strontium oven, capable of producing a $37\,\text{mrad}$ collimated atomic beam, with a flux of $1.0\times10^{13}\,\text{s}^{-1}\text{cm}^{-2}$ at the oven temperature of $450\,^{\circ}\text{C}$, reached with an electrical power consumption of $36\,$W. The latter value can be further reduced by employing a lower emissivity thermal shield ($\varepsilon<0.2$). Reduced dimension and high efficiency simplify the heat management of the vacuum system and keep the contribution of BBR frequency shift due to the oven on the strontium clock transition below the $10^{-18}$ fractional level. Laser-spectroscopy on the strontium transition $5\text{s}^2\,{}^1\text{S}_0\leftrightarrow5\text{s}5\text{p}\,{}^1\text{P}_1$ at $\lambda=461\,$nm was performed to characterize the atomic beam. We have shown that the collimation of the atomic beam allows frequency stabilization of a $461\,$nm laser on the atomic transition with a negligible systematic shift, in a simple linear absorption regime spectroscopy with a single probe beam. Atomic beam intensity was measured as a function of the oven temperature and was compared with the theoretical model. From the comparison we have concluded that the oven is operating in the collision-free regime. The simultaneous evidence that atomic beam intensity and divergence are respectively slightly below and beyond the theoretical values can be explained by an imperfect mutual alignment of the capillaries. From the theoretical model we have estimated a continuous operation lifetime of 10 years, experimentally we have not observed appreciable degradation of the atomic flux on a time scale of at least $2$ years. The measured peaking factor of the oven ($\chi\simeq8$) is compliant with a significantly simplified capillaries nozzle with $L/a\simeq20$. 

\begin{acknowledgments}
This work has been carried out in the framework of the project ``Space Optical Clocks" (SOC) funded within the ELIPS-3 program of the European Space Agency ESA, with co-funding by the German space agency DLR. The authors wish to acknowledge the support from the EU-7th framework programme SOC2 (grant no. 263500) and R. Ballerini, M. De Pas, M. Giuntini, A. Hajeb, A. Montori, E. Scarlini for technical assistance. 
\end{acknowledgments}

\providecommand{\noopsort}[1]{}\providecommand{\singleletter}[1]{#1}%

\end{document}